\documentstyle[12pt]{article}

\newcommand{\lsim}{\mathrel{\lower4pt\hbox{$\sim$}}
\hskip-12.5pt\raise1.6pt\hbox{$<$}\;}

\newcommand{\gsim}{\mathrel{\lower4pt\hbox{$\sim$}}
\hskip-12.5pt\raise1.6pt\hbox{$>$}\;}

\begin{document}
\baselineskip 22pt plus 2pt

\noindent \hspace*{10cm}SLAC-PUB-\\
\noindent \hspace*{10cm}TECHNION-PH-95-16\\
\noindent \hspace*{10cm}BNL-

\begin{center}
{\bf LARGE TREE LEVEL CP VIOLATION IN
$e^+e^-\to  t\bar{t}H^0$ IN THE TWO-HIGGS DOUBLET-MODEL}
\vspace{.7in}

 S. Bar-Shalom$^a$, D. Atwood$^b$, G. Eilam$^a$, R.R. Mendel$^{a,c}$ and A. Soni$^d$\\
\end{center}
\vspace{.2in}

\noindent
$^a$ Physics Dept., Technion-Israel Inst.\ of Tech., Haifa 32000, Israel.\\
$^b$ SLAC, P.O.Box 4349, Stanford, CA 94305, USA.\\
$^c$ On Sabbatical leave from: Dept.\ of Applied Math., Univ.\ of Western\\
\hspace*{0.30cm} Ontario, London, ON N6A 5B7, Canada.\\
$^d$ Physics Dept., Brookhaven Nat.\ Lab., Upton NY 11973, USA.
\vspace{0.5in}

\begin{center}
{\bf Abstract}\\
\end{center}

We find a large CP violation effect within the Two-Higgs-Doublet-Model for
the  reaction $e^+e^-\to  t\bar{t}H^0$ at future linear colliders.  The
CP-asymmetry arises already at the tree level as a result of interference
between diagrams with $H^0$ emission from $t$ (and $\bar{t}$) and its
emission from a $Z^0$ and can be about 10--20\%.  In the best case one needs
a  few  hundred 
$t\bar{t}H^0$ events to observe CP violation at the 3$\sigma$ level.

\pagebreak

Future high energy $e^+e^-$ colliders ($500\leq E_{cm}\leq 1000$ GeV) will
serve as a very useful laboratory for the study of Higgs and top physics
beyond the Standard Model (SM) \cite{zerwas}.  The top quark with a mass of
176 GeV \cite{abe},
being so heavy, is likely to be sensitive to the short distance physics  underlying
the SM above the electroweak scale. Searches for CP-violation in top physics
should be a  particularly useful probe of physics beyond the SM,
since it is unlikely that the CP violating KM phase \cite{kobayashi} in the SM can
account for the observed baryon asymmetry in the universe \cite{cohen}. One
of the simplest
extensions of the SM is the Two-Higgs-Doublet-Model (THDM) where one of the
two doublets is responsible for giving masses to the charge $+2/3$ quarks
and the other to the charge $-1/3$ quarks. This is also the
preferred supersymmetry-motivated THDM \cite{froggat,dawson}. We recall that CP
violation in top quark physics in such THDM has received considerable attention
in the past few years \cite{bernreuther}.

In this Letter we focus on CP violation, driven by THDM in the process $e^+e^-\to
t\bar{t}H^0$ at future $e^+e^-$ colliders, where $H^0$ is the lightest neutral
Higgs in the THDM\null. Within the SM, Higgs production
in $e^+e^-$ colliders was studied earlier at low \cite{gaemers} and high
\cite{djouadi} energies. It was also studied in the context of general THDMs,
in $Z$ decays \cite{kalinowski}, where recently CP violation in
Higgs  production within the THDM was examined in \cite{grzadkowski}.
We should emphasize that the
reaction we study is not meant (necessarily) to lead to the  discovery of
$H^0$ but rather to investigate the CP-properties of $H^0$. Clearly
even after the $H^0$ is discovered its role in CP violation will need to
be understood. This issue will thus be the main subject of our investigation.

A very interesting feature of the reaction $e^+e^-\to t\bar t H^0$ is that
it exhibits a CP asymmetry
 at the tree graph level. Such an effect arises from interference
of the Higgs emission from $t$ or $\bar t$ with the Higgs emission from the
$Z$ boson. Being a tree level effect the resulting asymmetry is quite large.
This asymmetry can be detected through
a CP-odd $T_N$-odd observable ($T_N$ is the naive time reversal operator
defined by replacing time with its negative without switching initial and final
states.)  In the best scenario one needs a few hundred $t\bar{t}H^0$
events to observe CP violation at the $3\sigma$ level.

In the THDM CP-violation may emanate from
the neutral Higgs sector.  In general, the manifestation of such CP-violation
is that the neutral Higgs mass eigenstates couple  to fermions
with both scalar and pseudoscalar couplings.

For $e^+e^-\to  t\bar{t}H^0$ the following
interaction terms in $\cal {L}$ are required \cite{froggat}:

\begin{eqnarray}
{\cal L}_{H_j^0} = H^0_j\bar{f}(a_{fj}+ib_{fj}\gamma_5)f +
H^0_j c_jg_{\mu\nu} Z^\mu Z^\nu +
\frac{c_j}{2M_Z} [\chi^0(\partial_\mu H^0_j) - (\partial_\mu\chi^0)H^0_j]
Z^\mu\ ,
\label{lhzj}%1
\end{eqnarray}

\noindent which involves the $f\bar{f}H^0_j$, $ZZH^0_j$ and $Z\chi^0H^0_j$
couplings. Here $f$ stands for a fermion, $\chi^0$ is the unphysical Goldstone boson
and $H_j^0$ is a neutral Higgs species. The three coupling constants, $a_{fj},
\ b_{fj}$ and $c_j$ are functions of tan$\beta$, which is the ratio
between the two vacuum expectation values in this model, i.e.\ tan$\beta = v_2/v_1$,
and of the three mixing angles, $\alpha_1, \ \alpha_2$ and $\alpha_3$
which diagonalize the Higgs mass matrix \cite{froggat}. In particular

\begin{eqnarray}
&&a_{fj} = - 2^{\frac{1}{4}} G_F^{\frac{1}{2}} m_f R_{2j}/\sin\beta \ \ , \ \ %2a
b_{fj} =
 - 2^{\frac{1}{4}} G_F^{\frac{1}{2}} m_f R_{3j}\cot\beta \ , \nonumber \\ %2b
&&c_j =
  2^{\frac{5}{4}} G_F^{\frac{1}{2}} M^2_Z(R_{j1}\cos\beta +
 R_{j2}\sin\beta) \ . \label{cj} %2c
\end{eqnarray}

\noindent $R$ is the rotation matrix given by:

\setcounter{equation}{2}
\def\theequation{\arabic{equation}}
\begin{eqnarray}
R=\left(\matrix{ & c_1 & s_1c_3  & s_1s_3 \cr
&-s_1c_2 & c_1c_2c_3-s_2s_3 & c_1c_2s_3 + s_2c_3 \cr
&s_1s_2 & -c_1s_2c_3-c_2s_3 & -c_1s_2s_3 + c_2c_3 } \right) \ , \label{req}
\end{eqnarray}

\noindent where $s_i\equiv \sin \alpha_i$ and $c_i\equiv \cos \alpha_i$.

We now discuss the tree-level cross-section and CP-violation effects in our
reaction,

\begin{eqnarray}
e^+(p_+)+e^-(p_-)\to t(p_t) + \bar{t}(p_{\bar{t}}) + H^0(p_{H}) \ . \label{eppp}
%6
\end{eqnarray}

\noindent We assume that two of the three neutral Higgs particles
are much heavier than the remaining one, i.e.\ $H^0$.
We therefore omit the index $j$ in Eqs.~\ref{lhzj} and \ref{cj}, and denote
the couplings as: $a_t$, $b_t$ and $c$.
An important property of this simple
reaction, is that it gives rise to CP-violation
already at tree-level, as a result of interference of the diagram with $H^0$ emitted from
the $Z$ with the diagram where $H^0$ is radiated off the
$t$ or $\bar{t}$. The tree-level differential
cross section $\Sigma^0$ is a sum of two terms: the CP-even and odd terms $\Sigma^0_+$ and
$\Sigma_-^0$, respectively, i.e.\ $\Sigma^0\equiv \Sigma_+^0 + \Sigma^0_-$.
$\Sigma^0_\pm$ are calculated from the tree-level diagrams in Fig.~1.

The incoming left or right polarized electron-positron current can be written as:

\begin{eqnarray}
J_e^{\mu(j)} = \bar{v}_e(p_+)\gamma^\mu \frac{1+j\gamma_5}{2} u_e(p_-) \
,    \label{jmue}     %8
\end{eqnarray}

\noindent where  $j=-1(1)$ for left(right) handed electrons. We write the
tree-level amplitude as:

\begin{eqnarray}
{\cal M}^0 = \sum_\alpha \sum_\rho \mu^\alpha_\rho \ ,  \label{mzero} %10
\end{eqnarray}

\noindent where $\rho$ indicates the diagram ($\rho=i,ii,iii$ for diagrams
$i,ii,iii$, respectively in Fig.~1) and
$\alpha$ indicates the gauge particle exchanged in the $s$-channel, i.e.\ $\alpha=Z, \gamma$.
We then write the general form of $\mu^\alpha_\rho$ as:

\begin{eqnarray}
\mu^\alpha_\rho = J_e^{\mu(j)} \bar{u}(p_t)H^\alpha_{\rho\mu} v(p_{\bar{t}})
\ ,  \label{mualpha} %11
\end{eqnarray}

\noindent where $H^\alpha_{\rho\mu}$,  corresponding to each diagram, are
given by:

\begin{eqnarray}
&& H^Z_{i\mu} = - C_Z \pi_t(a_t+ib_t\gamma_5)(p\hspace{-0.20cm}/_t
+ p\hspace{-0.20cm}/_H
+ m_t)\gamma_\mu C_{LR}^+ \ , \label{hzi}      \\ %12a
&& H^\gamma_{i\mu} = H^Z_{i\mu}(C_Z\to -C_\gamma, \ C^+_{LR}\rightarrow
1) \ ,     \label{hgammai}     \\     %12b
&&H^Z_{ii\mu} = C_Z\pi_{\bar{t}}\gamma_\mu C^+_{LR}
(p\hspace{-0.20cm}/_{\bar{t}} + p\hspace{-0.20cm}/_H - m_t)(a_t+ib_t\gamma_5)
\ ,  \label{hzii} \\ %12c
&& H^\gamma_{ii\mu} = H^Z_{ii\mu} (C_Z\to -C_\gamma, C^+_{LR}\to
1) \ ,  \label{hgammaii}     \\                      %12d
&& H^{ZZ}_{iii\mu} = C_Z\pi_{ZH} c\gamma_\mu C^+_{LR} \ ,  \label{hzz}    \\ %12e
&& H^{Z\chi^0}_{iii\mu} = C_Z \pi_{ZH} c(c^t_R-c^t_L)m_t\gamma_5
p_{H^0\mu}/ m^2_Z
\ , \label{hzchi}  %12f
\end{eqnarray}

\noindent where $C_Z \equiv \pi\alpha\pi_zc^e_j/c_W^2s_W^2$,
$C_\gamma \equiv 4\pi\alpha Q_q\pi_\gamma$.
$Q_q$ is the charge of the quark in the final state and $c_W(s_W)$ stands for
cos$\theta_W(\sin\theta_W)$.  $c^e_j = c^e_L(c^e_R)$ for
$j=-1(1)$ where $c^f_L = - 2 I_3^f + 2Q_fs_W^2$ and $c^f_R = 2Q_fs_W^2$.
$\pi_Z$ and $\pi_\gamma$ are the $Z$ boson and photon propagators,
respectively and:

\begin{eqnarray}
&&\pi_t\equiv \frac{1}{2p_t\cdot p_{H^0}+m^2_{H^0}} \ , \
\pi_{\bar{t}}\equiv \frac{1}{2p_{\bar{t}}\cdot p_{H^0}+m^2_{H^0}} \nonumber \\
&&\pi_{ZH}\equiv \frac{1}{P^2 - 2 P\cdot p_{H^0}+m^2_{H^0}-
m^2_Z} \ . \label{pit}
\end{eqnarray}

\noindent Furthermore, $P \equiv p_-+p_+ \ {\rm and} \
C^+_{LR} \equiv c^t_L L + c^t_R R$, $L,R = (1\mp \gamma_5)/2$.
With the above definitions $\Sigma^0$ is given by:

\begin{eqnarray}
\Sigma^0 = \frac{1}{2}\sum_j|\sum_\alpha\sum_\rho\mu^\alpha_\rho|^2 \ , \label
{sigmaz} %13
\end{eqnarray}

\noindent where the sum over $j$ is carried over the polarization of $e^+,
t$  and $\bar{t}$. Also $\rho=\{i,ii,iii\}$ and $\alpha=Z,\gamma$.
The factor of $1\over 2$ is due to the fact that we consider an unpolarized 
$e^+$ beam colliding with a polarized $e^-$ beam. The expression for  
$\Sigma^0_+$ is quite involved and will not be given explicitly here.

For illustration, we adopt the value tan$\beta=0.5$ which
gives large effects and is allowed by present experiments
for $m_{H^+}\geq \cal{O}$(400 GeV), where $H^+$ is the charged Higgs
boson of the THDM  \cite{hewett,gunion}.
We plot the tree-level cross-section for
$e^+e^-\to t\bar{t}H^0$ in Fig.~2 for $m_{H^0} = 100$ and
160 GeV, for two possible sets of the Higgs coupling
constants $a_t, \ b_t$ and $c$. Set I corresponds to tan$\beta = 0.5, \ \alpha_1=
\pi/4$,  $\alpha_2 = \pi/4$, $\alpha_3=0$ and
set II - to: tan$\beta = 0.5, \ \alpha_1=\pi/4$,  $\alpha_2 =
\pi/2$, $\alpha_3=0$. The CP-violating piece of the tree-level differential
cross-section is:

\begin{eqnarray}
\Sigma^0_- &=& 2 C_Z\frac{m_t}{M_Z^2}\pi_{ZH}
cb_t \times E \times \left\{  \; j \times (\pi_t+\pi_{\bar{t}}) \right. \nonumber \\ &\times&\left[
(s-s_t-M_H^2)
(C_Z(c_R^t+c^t_L)-2C_{\gamma}) - 4 C_{\gamma} (c_R^t-c^t_L) M_Z^2 \right]  \nonumber \\
&+&  \left. 2 C_Z f (c_R^t-c^t_L) (\pi_t-\pi_{\bar{t}}) \right. \left. \right\}
\ , \label{sigmazm}
\end{eqnarray}

\noindent where: $E \equiv \epsilon(p_-,p_+,p_t,p_{\bar{t}})$, $s\equiv 2p_-\cdot
p_+$, $s_t\equiv(p_t+p_{\bar{t}})^2$, $f\equiv (p_- - p_+)\cdot(p_t+p_{\bar{t}})$
and $j = - 1(1)$ for a left(right) handed electron.

Of course, at tree-level there are no absorptive phases. Thus the CP-violating
term $\Sigma^0_-$ can probe only CP-asymmetries of the $T_N$-odd type.
This leads us to consider the following CP-odd,
$T_N$-odd, triple correlation product

\begin{eqnarray}
O = \vec{p}_-\cdot(\vec{p}_t\times\vec{p}_{\bar{t}})/s^{3/2} \ . \label{oeq} %15
\end{eqnarray}

\noindent To observe a non-vanishing average value $\langle O \rangle$ with
a  statistical significance of $\sigma$ in an ideal experiment, one needs:

\begin{eqnarray}
N_{t\bar{t}H^0}^\sigma = \sigma^2/A^2_O \ ,       \label{nsigma}        %16
\end{eqnarray}

\noindent events, where $A_O$ is given by:

\begin{eqnarray}
A_O \equiv \langle O \rangle/\sqrt{\langle O^2\rangle} \ .  \label{az} %17
\end{eqnarray}

\noindent The number of expected $t\bar{t}H^0$ events is
$N_{t\bar{t}H^0} =
{\cal L} \times \sigma (e^+e^-\to t\bar{t}H^0)$ where
${\cal L}$ is the collider luminosity.  Fig.~3 shows our main
results for $m_H=100$ and 160 GeV, for set II of
tan$\beta,\alpha_1,\alpha_2$ and $\alpha_3$.
We have also used $m_t=176$ GeV \cite{abe} and took the electron to be unpolarized.
We have depicted the number of events, $N_O$, required to observe a non-vanishing
value (to one sigma) for the $T_N$-odd observable $\langle O \rangle$.
We have also plotted in Fig.~3 the expected number of
$t\bar{t}H^0$ events per year, $N_{\exp}$, in an $e^+e^-$ linear collider with a luminosity of
${\cal L} = 3 \times 10^{33}$cm$^{-2}$sec$^{-1}$ for CM energies
of $\sqrt{s} \cong 500-1000$ GeV\null. We see that near threshold, at $\sqrt{s}
\cong 500$ GeV, CP-violation asymmetry is far too small to be observed, i.e.\ $N_O\gg
N_{\rm exp}$. However, in this scenario
$N_O$ and $N_{\exp}$ do cross each other.  For $m_{H^0}=100$ GeV the crossing
appears at $\sqrt{s}\cong 800$ GeV and for $m_{H^0}=160$ GeV at $\sqrt{s}\cong850$
GeV\null. This crossing means that for set II of the parameters, one may be
able to observe CP-violation (to one sigma) in the process $e^+e^-\to t \bar{t}
H^0$, at CM energies of $\sqrt{s}\cong800-1000$ GeV and for Higgs masses of
100--160 GeV\null.  For example, for $m_{H^0}=100$ GeV and at $\sqrt{s}\cong1000$
GeV,  $N_O/N_{\exp} \cong 0.65$.
The results for the set I of parameters do not look as promising.
For example, for $\sqrt{s}\cong1000$ GeV and $m_{H^0}=100$ or 160 GeV we obtain
$N_O/N_{\exp}\cong4$. Typically, at least several hundreds of events
are needed in this case for a $1\sigma$ effect, as opposed to tens of events for set II.

%%%%%%%%%%%%%%%%%%%%%%%%%%%%%%%%%%%%DA
%
%
%
%
%
%

In Fig.~4 we show the dependence of the ratio $N_O/N_{\exp}$
on $\tan\beta$ for $m_H=100$ and 160 GeV and for $\sqrt{s}=800$ and
1000 GeV\null. We have kept all other parameters the same as in Fig.~3.
We see that $N_O/N_{\exp}$ depends only mildly on $\tan\beta$ for
$0.2\lsim \tan\beta\lsim1$.

For a given model of CP-violation it can be shown \cite{atwood} that
the  optimal observable to use is given by:

\begin{eqnarray}
O_{\rm iopt} = \frac{\Sigma^{Im}_-}{\Sigma^0_+} \ \ , \ \
O_{\rm ropt} = \frac{\Sigma^{Re}_-}{\Sigma^0_+} \ ,  \label{oiopt} %18
\end{eqnarray}

\noindent where the superscripts $Im$ and $Re$ refer to that part of the amplitude
proportional to the $\sin$ or $\cos$ of an absorbtive phase. Since absorbtive
effects require at least 1-loop, the $T_N$-even asymmetry such as for $O_{\rm
iopt}$ are smaller  by  a factor of order $\alpha_s/\pi$.

In Table~\ref{oropt}  we present our  results (for set II) for the
number of events required to detect a non-vanishing $\langle O
\rangle$ and $\langle O_{\rm ropt} \rangle$ (to one sigma) for
3 different CM energies, with $m_{H^0}=100$ or 160 GeV.
To illustrate the effect of polarization, we have included
in the table results for $O_{\rm ropt}$ and $O$ for different
polarizations of the electron. Of course,
$O_{\rm ropt}$ is related to $O$ by multiplication by a 
CP-even function since there is only one possible independent triple product
correlation when the final-state consists of three particles only.
As can be seen from the numbers,  the CP asymmetries (see eqns.
18 and 19), are in the 10--20\% range and $O$ gives almost 
as good results as
$O_{\rm ropt}$.

Note that to be able to measure the above observables, one would have to
reconstruct the transverse components of the $t$ and the 
$\bar{t}$ momenta in each $t\bar{t}H^0$ event.
This may be difficult in practice. Therefore, we present results for
an additional observable which 
requires only the determination of the momenta of the $b$ and
$\bar{b}$ in the process  $e^+e^-\to  t\bar{t}H^0 \to bW^+\bar{b}W^-H^0$.
Define the observable:

\begin{eqnarray}
O_b = \epsilon(p_-,p_+,p_b,p_{\bar{b}})/s^2 \label{ob}
\end{eqnarray}

\noindent In Table~\ref{osubb} we present our results for $O_b$. It is evident
that the number of events required to observe a $1\sigma$ CP-violating 
effect is comparable to the numbers found  with the observable $O$, in particular, 
for $\sqrt{s}\cong800-1000$ GeV\null.  Close to threshold, e.g.\ $\sqrt{s}
\cong600$ GeV, the effect would be much harder to observe through $O_b$.

There are two other comments that we wish to make in brief. First we
note that the $t\bar t H^0$ final state is expected to be the focus of
intense scrutiny to unravel in detail the interaction of the Higgs with
the top quark. Thus it is especially gratifying that the promising
signal for CP violation that our study indicates are expected in the
same final state. We note also that the method seems most suitable for
a Higgs of mass $\lsim160$ GeV\null. Such a Higgs will decay
predominantly into $b\bar b$ with a $BR$ of $O(1)$.

To summarize, CP-violation in Higgs emission at a future high energy $e^+e^-$
collider was investigated within the THDM\null. In the SM such a CP asymmetry
will vanish at least to two loop orders in perturbation theory and therefore
is expected to be extremely small.  In contrast, in the THDM, an important
and very interesting property of the reaction
$e^+e^-\to t\bar{t}H^0$ is that the CP-violation arises already at tree-level
through interference of $H^0$ emission from $t$ or $\bar{t}$ and its emission
off a $Z$-boson, and therefore allows for large CP-violation effects. It
is clearly important to examine this effect in other extensions of the SM\null.
Our main
result is that the CP-asymmetry in this process could be observable 
at a future linear $e^+e^-$ collider with a luminosity of the order
${\cal L} \approx 10^{33}$ cm$^{-2}$sec$^{-1}$ running at CM energies 
around 800-1000 GeV\null. We also showed that the
reaction is quite promising for $0.2\lsim \tan\beta\lsim1$ although
$\tan\beta\simeq 0.5$ seems to be the most suitable.

\bigskip
\bigskip

The work of RRM was supported in part by the Natural Sciences and
Engineering Research Council of Canada. SBS wants to thank Y. Ben-Horin
for his helpful advice with regard to the computer programs for numerical
evaluation of the results.  The research of G.E. has been supported 
in part by the BSF and by the Fund for the Promotion of Research at the Technion.
The work of DA was supported by US Department of Energy contract
DE-AC03-765F00515 (SLAC)  while the work of AS was supported by
US Department of Energy contract DE-AC02-76CH0016 (BNL).
\pagebreak

%\pagebreak

\begin{center}
{\bf Figure Captions}
\end{center}

\begin{description}

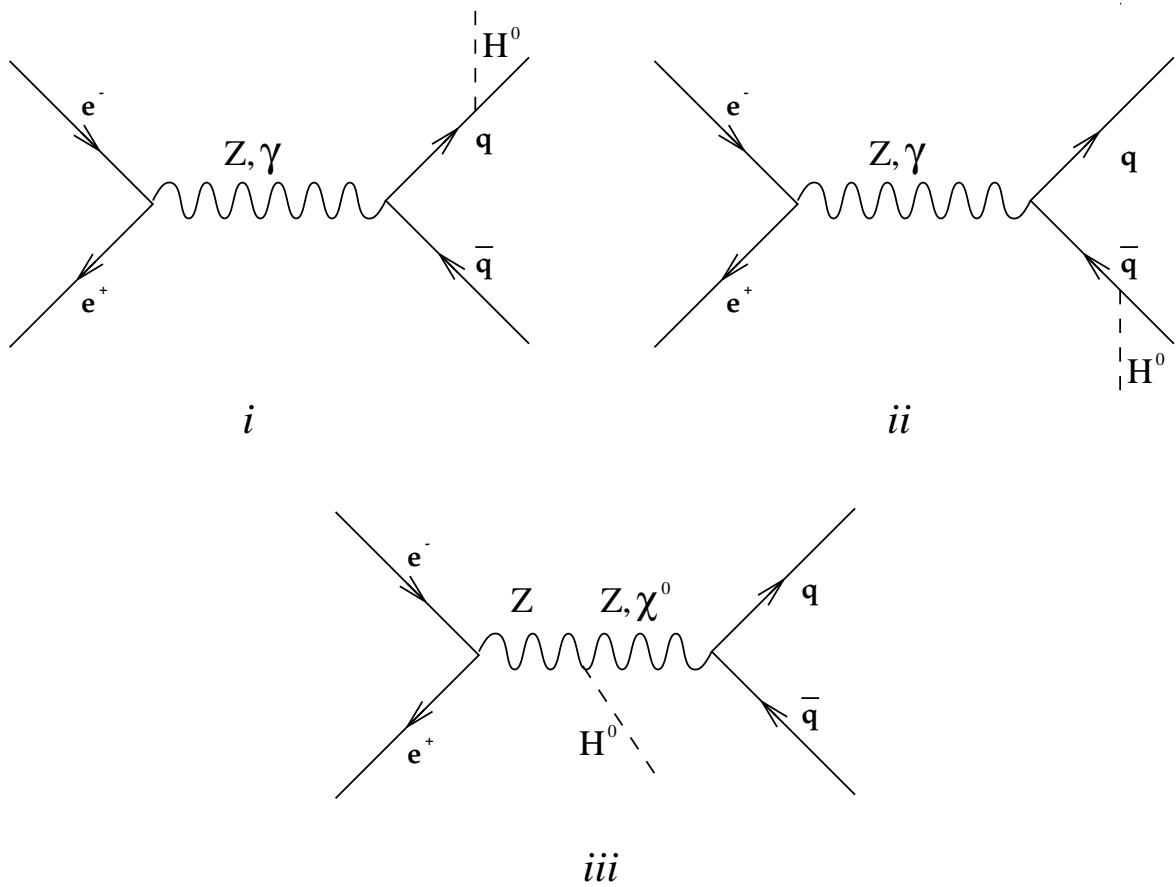
\item{Fig. 1:} Tree-level Feynman diagrams contributing to
$e^+e^-\to t\bar{t}H^0$ within the two Higgs doublet model.

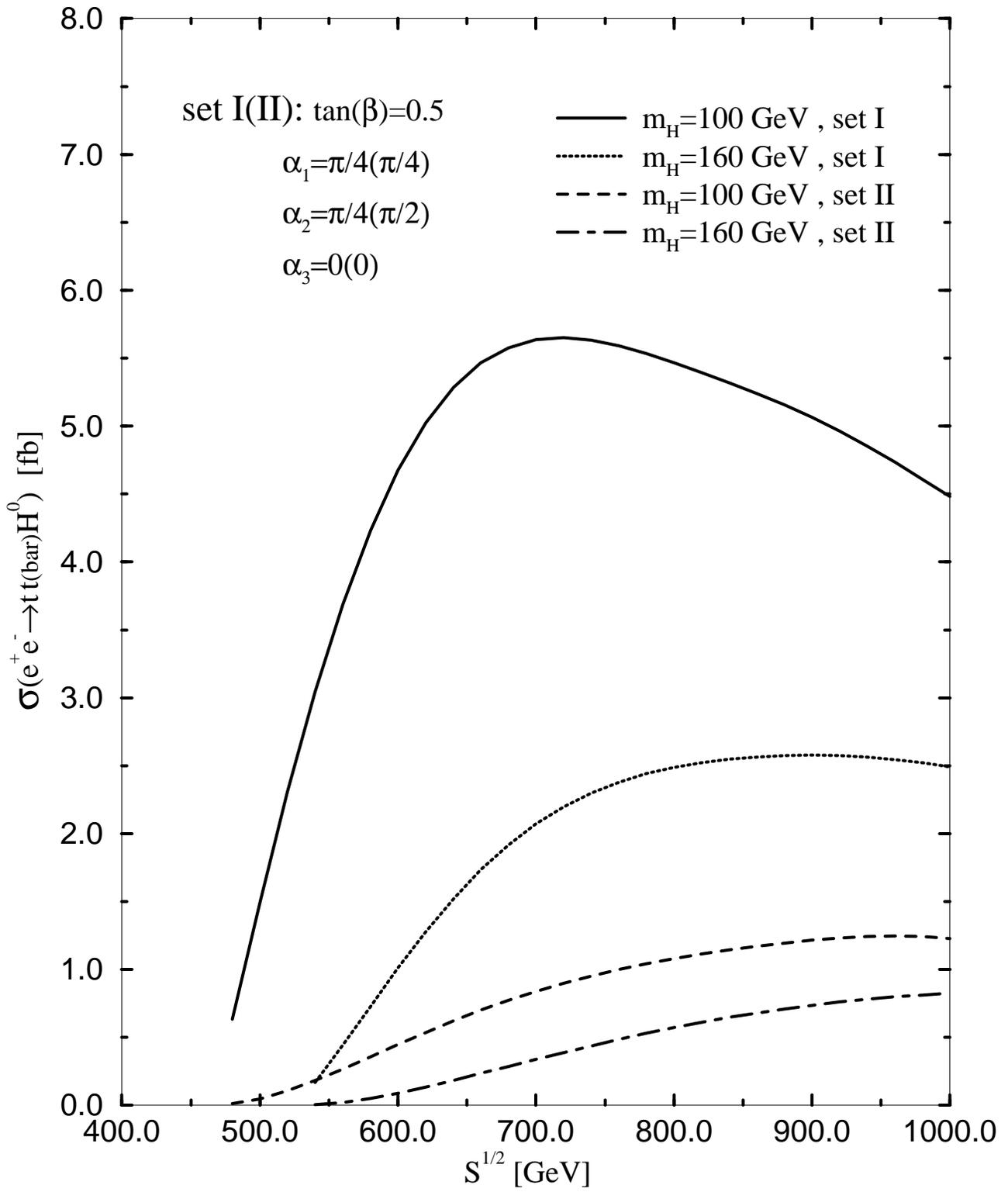
\item{Fig. 2:} The cross section for the reaction
$e^+e^-\to t\bar{t}H^0$, for sets I and II of the parameters
 $a_t, \ b_t$ and $c$ and for $m_{H^0}=100$ and 160 GeV
assuming unpolarized electron and positron beams.

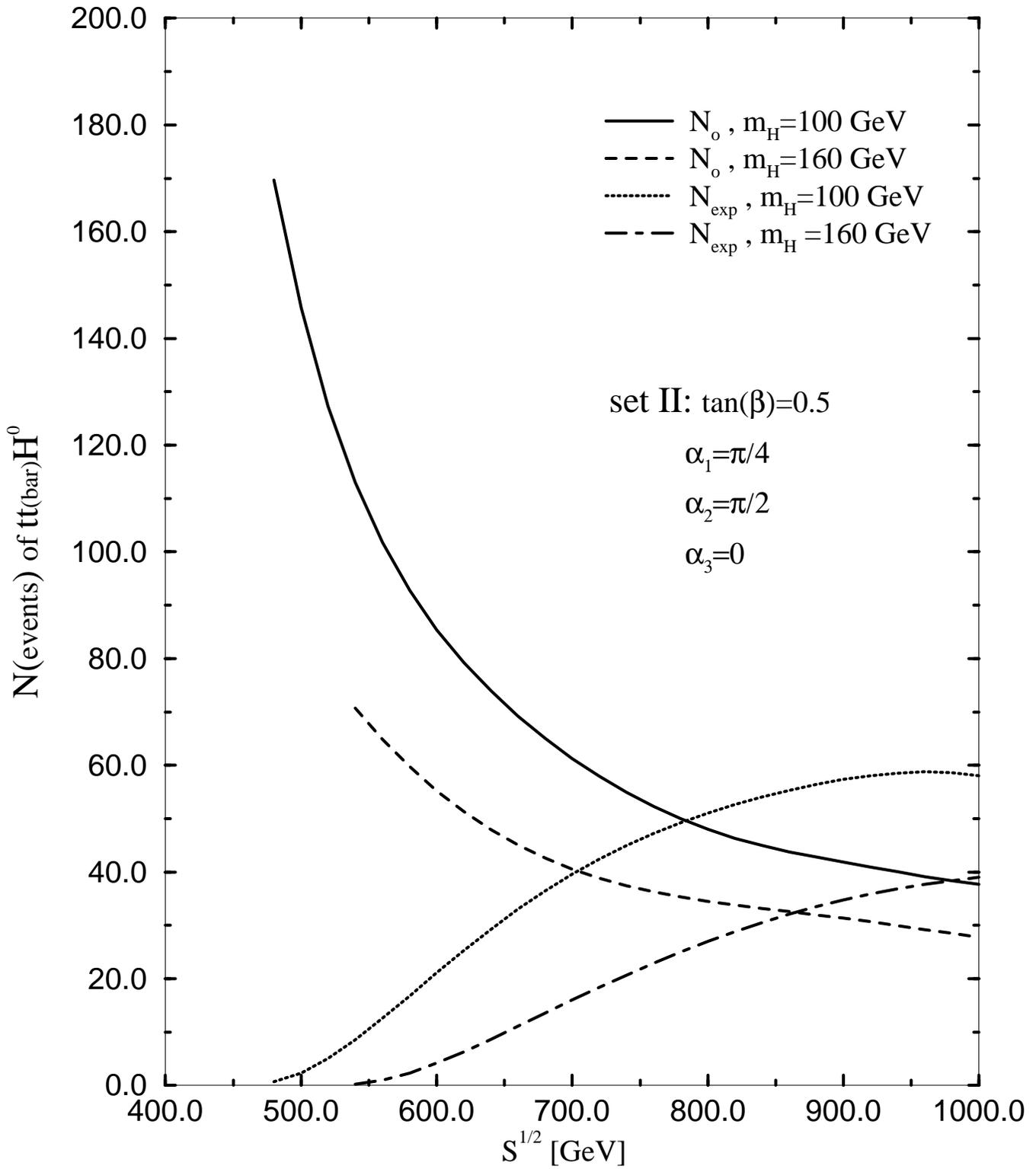
\item{Fig. 3:} Number of events, $N_O$, required to detect CP-violation via
$\langle O \rangle$ at $1\sigma$ 
level and the expected yearly number of events $N_{\exp}$,
as a function of total beam energy for set II 
of the parameters and for $m_{H^0}=100$ and 160 GeV
with unpolarized electron and positron beams.

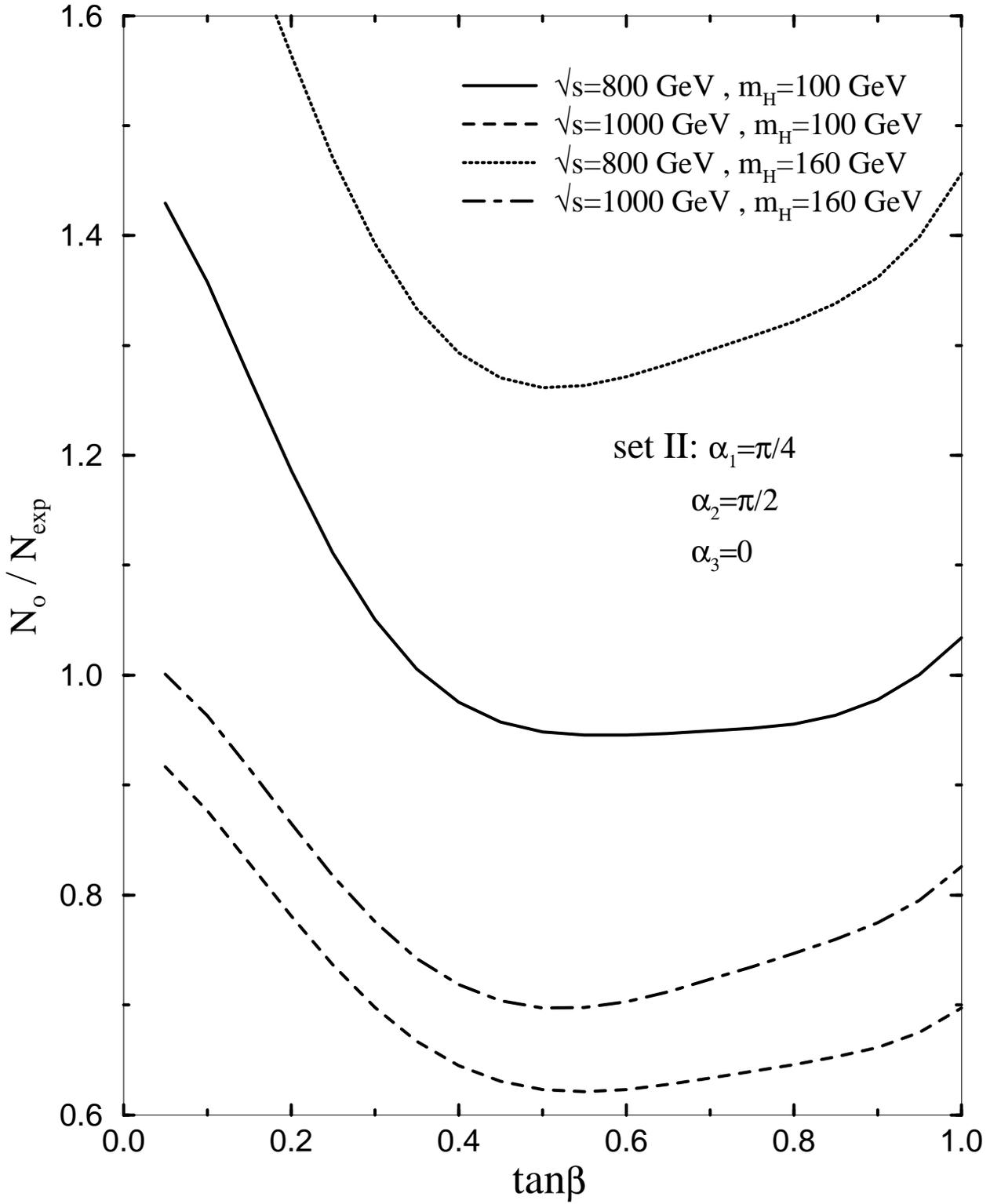
\item{Fig. 4:} $N_O/N_{\rm exp}$ versus $\tan\beta$ for $m_{H^0}=100$
and 160 GeV and $\sqrt{s}=800$ and 1000 GeV\null. The other parameters
are held the same as in the previous figures.
\end{description}

\newpage

\begin{table}
\begin{center}
\caption[first entry]{The number of events needed to detect $\langle O
\rangle$ and $\langle O_{\rm ropt} \rangle$ at $1\sigma$
is given for sets II
of the parameters $a_t, \ b_t$ and $c$.  The left and right
polarization ($j=-1$ and 1, respectively) is compared with the unpolarized
case. The values of $\sqrt{s}$ and $m_{H^0}$ are given in GeV\null.
\protect\label{oropt}}
\begin{tabular}{|r|r||r|r|r|r|} \cline{3-6}
\multicolumn{2}{c||}{~~} & \multicolumn{4}{c|}{Set II}\\ \hline
$\sqrt{s}$ & j
& \multicolumn{2}{c|}{$O$}
& \multicolumn{2}{c|}{$O_{\rm ropt}$}
\\ \cline{3-6}
& & $m_{H^0}=100$ & $m_{H^0}=160$ &$m_{H^0}=100$ & $m_{H^0} = 160$\\ 
\hline
\hline
&-1& $100$ & $70$ & 95 & 65\\ \cline{2-6}
600 & unpol & $85$ & $55$ & 80 & 55 \\ \cline{2-6}
& 1 &  $60$ & $40$ & 60 & 40 \\ \hline
\hline
& -1 &  $60$ & $40$ & 50 & 40 \\ \cline{2-6}
800 & unpol & $50$
& $35$ & 45 & 35 \\ \cline{2-6}
&1 &  $40$ & $25$
& 35 & 25 \\ \hline \hline
& -1 &  $40$ & $35$ & 35 & 30 \\ \cline{2-6}
1000 & unpol &  $40$ & $30$ & 30 & 25 \\ \cline{2-6}
& 1 &  $30$ & $25$ & 25 & 20 \\ \hline
\end{tabular}
\end{center}
\end{table}
\vspace{0.75cm}

\begin{table}
\begin{center}
\caption{The same as table 2 except for $\langle O_b \rangle$. \label{osubb}}
\begin{tabular}{|r|r||r|r|} \cline{3-4}
\multicolumn{2}{c||}{~~} & \multicolumn{2}{c|}{Set II}\\ \hline
$\sqrt{s}$ & j
& \multicolumn{2}{c|}{$O_b$}
\\ \cline{3-4}
& & $m_{H^0}=100$ & $m_{H^0}=160$ \\ 
\hline
\hline
& -1 & $185$ & $180$ \\  \cline{2-4}
600 & unpol & $205$ & $270$ \\  \cline{2-4}
& 1 & $275$ & $1280$ \\ \hline
\hline
& -1 & $70$ & $50$ \\ \cline{2-4}
800 & unpol & $65$ & $45$ \\ \cline{2-4}
&1 & $55$ & $40$ \\ \hline \hline
& -1 & $50$ & $35$ \\ \cline{2-4}
1000 & unpol & $45$ & $30$ \\ \cline{2-4}
& 1 & $35$ & $25$ \\ \hline \hline
\end{tabular}
\end{center}
\end{table}

\end{document}